\documentclass[12pt]{article}

\usepackage{graphicx}

\newcommand{\ins}{\mbox{s$^{-1}$}}
\newcommand{\RS}{\mbox{$\Lambda_S$}}
\newcommand{\RT}{\mbox{$\Lambda_T$}}
\newcommand{\RD}{\mbox{$\Lambda_D$}}
\newcommand{\RQ}{\mbox{$\Lambda_Q$}}
\newcommand{\qr}{\mbox{$\lambda_q$}}
\newcommand{\dr}{\mbox{$\lambda_d$}}
\newcommand{\qdr}{\mbox{$\lambda_{qd}$}}

\begin{document}

\begin{center}
{\Large\bf \boldmath Nuclear Muon Capture in Hydrogen and its Interplay with Muon Atomic Physics}

\vspace*{6mm}
{Peter Kammel$^{1,2,}$\footnotemark[3]\footnotetext[3]{Work supported by U.S. 
National Science Foundation, Department of Energy, NCSA and CRDF, Paul Scherrer Institute, the Russian Academy of Sciences and the Russian Federation.}}\\
{\small \it $^1$ University of Illinois at Urbana-Champaign, Urbana, Illinois 61801, USA \\ 
            $^2$ representing the MuCap Collaboration~\cite{MuCap}}
\end{center}

\vspace*{6mm}

\begin{abstract}
The singlet capture rate \RS\ for the semileptonic weak 
process $\mu+p \rightarrow n+\nu_\mu$ has been measured in the MuCap experiment~\cite{Andreev:2007wg}. The
novel experimental technique is based on stopping muons in an active target, consisting of a
time projection chamber operating with ultra-pure hydrogen. This allows the unambiguous 
determination of the pseudoscalar form factor $g_P$ of the charged electroweak current 
of the nucleon. Our first result \mbox{$g_P(q^2=-0.88 m^2_\mu) = 7.3 \pm 1.1 $} is 
consistent with accurate theoretical predictions and constitutes an important test of QCD symmetries. 
Additional data are being collected with the aim of a three-fold reduction of the experimental 
uncertainties. Building on the developed advanced techniques,
the new MuSun experiment~\cite{MuSun} is being planned to measure the muon capture rate on the 
deuteron to 1.5\% precision. This would provide  the  by far most accurate experimental information 
on the axial current interacting with the two-nucleon system and determine the low energy constant 
$L_{1A}$ relevant 
for solar neutrino reactions.

Muon induced atomic and molecular processes represent challenges as well as opportunities for this 
science program, and their interplay with the main nuclear and weak-interaction physics aspects 
will be discussed. 
\end{abstract}

\vspace*{6mm}

\section{Muon Capture on Hydrogen Isotopes}

The processes

\begin{equation}
\mu + p \rightarrow n + \nu_\mu     
\label{m+p.eq}
\end{equation}
\begin{equation} 
\mu + d \rightarrow n + n + \nu_\mu 
\label{m+d.eq} 
\end{equation}
are fundamental weak reactions between a muon and the nucleon (\ref{m+p.eq}) and
the simplest nucleus (\ref{m+d.eq}), respectively. Historically, they have played an
important role in establishing the helicity structure as well as the universality of the weak
interaction. This paper is dedicated to the pioneers of this era. E. Zavattini, who
inspired us with his experiments and enthusiasm for this research direction and 
V.P. Dzhelepov and V.G. Zinov, whose leadership stimulated many key experiments in muon 
capture and muon-catalyzed fusion at the JINR, Dubna. 

Today, the electro-weak interaction is understood and verified with amazing precision at the
quark-lepton level. Thus, the charged lepton current 
serves as a clean probe for exploring the weak couplings and QCD structure of the nucleon and nuclei. Over the last
decade the connection between nucleon and even two-nucleon observables at low energies 
and fundamental QCD has been elucidated  by the development of modern
effective field theories (EFT). In this framework, process  (\ref{m+p.eq}) can be calculated 
in a model-independent way with controlled systematic uncertainty. As the EFT predictions 
derive from basic concepts of explicit and spontaneous chiral symmetry breaking, their experimental 
verification is an important test of our understanding of the underlying QCD symmetries. As 
regards the two-nucleon sector, EFT calculations have proved that reaction (\ref{m+d.eq}) 
is closely related to fundamental weak reactions of astrophysical interest, like $pp$ fusion 
in the sun and $\nu d$ scattering observed at the Sudbury Neutrino 
Observatory. A precision measurement of reaction (\ref{m+d.eq}) comes closest to calibrating
these extremely feeble reactions under terrestrial conditions.    

\begin{table}[htb]
\begin{center}
\begin{tabular}{ccccc}
\hline 
atom	& $\Delta E_{hfs}$(meV) & hfs state & capture rate & theory value(\ins) \\
\hline 
$\mu p$	&   182.0 		& triplet (F=1) 		& \RT 		&  13.3\cite{Bernard:2000et}   \\  
        &                       & singlet (F=0)			& \RS		&  706.6\cite{Bernard:2000et}, 714.5\cite{Ando:2000zw}, 711.5\cite{Czarnecki:2007th}	\\
$\mu d$	&    48.5 		& quartet (F=$\frac{3}{2}$) 	& \RQ 		&  $\approx$12.5   \\  
        &                       & doublet (F=$\frac{1}{2}$)	& \RD		&  386\cite{Ando:2001es}	\\
\hline 
\end{tabular}
\caption{Hyperfine energy splittings and recent calculations of capture rates from hfs states 
of $\mu p$(1S) and $\mu d$(1S) atoms. Theoretical predictions~\cite{Bernard:2000et, Ando:2000zw} 
updated with recently calculated radiative corrections~\cite{Czarnecki:2007th}.   }
\label{rates.tab}
\end{center}
\vspace{-.5cm}
\end{table}

It has been realized early on, that the weak muon capture reactions and 
muon-atomic and molecular processes (muon kinetics) are closely intertwined. 
This is natural, as muon capture takes place not in flight, but from bound muonic hydrogen 
atoms and even molecules. Most notably, the V-A structure of weak interactions favors
capture from the lower hyperfine state of the muonic hydrogen atom (see Table~\ref{rates.tab}), 
so that the experimentally observed capture rate is largely proportional to the population of these states. 
Exchange collisions depopulate the initial statistical hyperfine population towards these
energetically lower lying states, the singlet $\mu p$ and doublet $\mu d$, respectively, whereas  
after $pp\mu$ formation, the hyperfine populations change again, due to the comparatively slow
conversion between the ortho- and para state of this muonic molecule. 
In the past, this complexity of muon atomic physics has seriously challenged the extraction of 
reliable weak interaction information from muon capture in hydrogen.

The new generation of precision muon capture experiments are characterized by novel experimental 
techniques and utmost care in eliminating ambiguities due to kinetic effects. This effort benefits
from the tremendous knowledge gained in the course of muon-catalyzed fusion research. 
We will discuss the MuCap experiment on reaction (\ref{m+p.eq})
and its first results, present the new MuSun experiment on reaction (\ref{m+d.eq}) and summarize
the closely related measurements of the positive muon lifetime. 


\section{The MuCap Experiment}
\label{MuCap}

Muon capture on the proton (\ref{m+p.eq}) is 
described by the low energy current-current form of weak interaction, where the vector- and axial vector 
matrix elements of the nucleon current are given by
\begin{eqnarray}
V_\alpha &=& \bar{u}_n(g_v(q^2) \gamma_\alpha         + \frac{i g_m(q^2)}{2 M_N} \sigma_{\alpha\beta} q^\beta) u_p \\
A_\alpha &=& \bar{u}_n(g_a(q^2) \gamma_\alpha\gamma_5 + \frac{g_p(q^2)}{m_\mu} q_\alpha \gamma_5)  u_p
\end{eqnarray}
in the most general form compatible with Standard Model symmetries. For the relevant moderate $q^2_0=-0.88\,m_\mu^2$,
form factors ${\textsl g}^{}_V(q^2_0)$, ${\textsl g}^{}_M(q^2_0)$ and ${\textsl g}^{}_A(q^2_0)$ are 
accurately determined  and contribute an uncertainty of only 0.46\% to $\Lambda^{}_S$. 
Process~(\ref{m+p.eq}) provides the most direct probe of 
${\textsl g}^{}_P \equiv {\textsl g}^{}_P(q^2_0)$, the pseudoscalar 
coupling of the nucleon's axial current, which is, experimentally, by far the 
least well known of these form factors. In EFT (chiral perturbation theory, ChPT), however,
$g_P$ is a derived quantity, which has been systematically evaluated up
to two-loop order~\cite{Kaiser:2003dr} and precisely calculated
${\textsl g}^{}_P=8.26\pm0.23$~\cite{Bernard:2001rs}. Efforts to calculate
$g_P$ on the lattice are progressing~\cite{Alexandrou:2007zz,Lin:2008uz}. The
experimental verification of these predictions is an important test of QCD at low 
energies~\cite{Bernard:2001rs,Gorringe:2002xx,Govaerts:2000ps}. 
\begin{figure}[hbt] 
\vspace{-.0cm}  
\begin{center}
\begin{tabular}{|c|} \hline
\includegraphics[scale=0.45, angle=0]{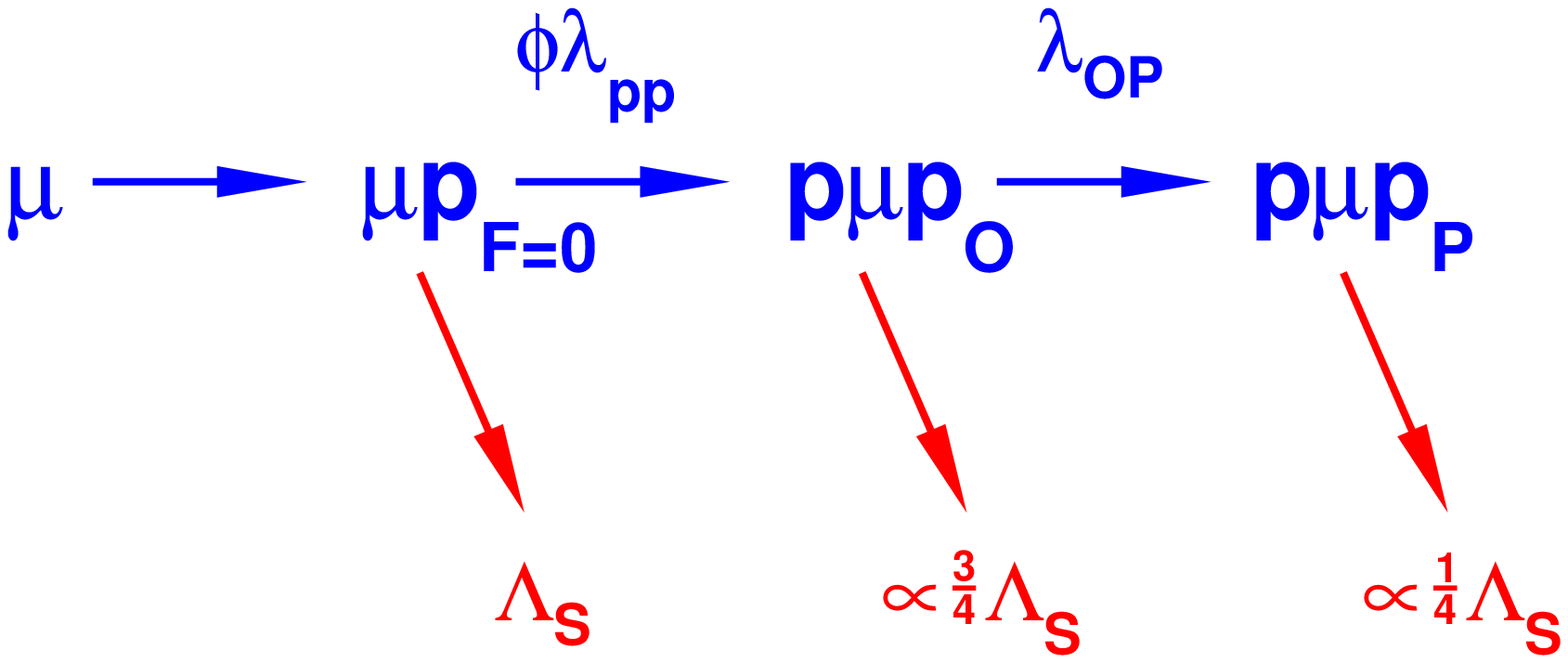} \\
\hline 
\includegraphics[scale=0.45, angle=0]{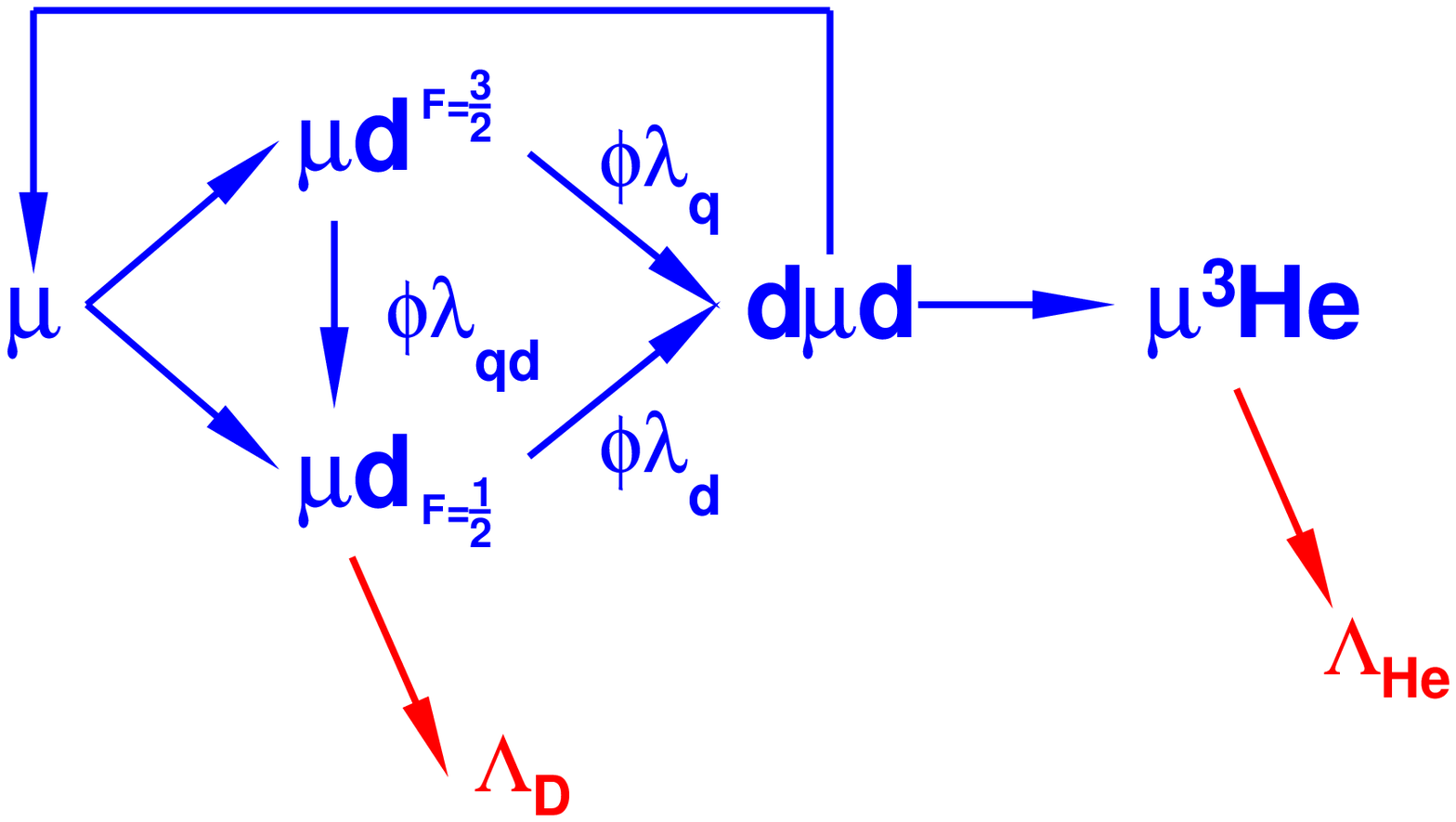} \\
\hline 
\end{tabular}
\vspace{-.0cm}
\caption{\label{kin.fig} Simplified scheme of muon induced reactions in pure hydrogen (top) and deuterium (bottom). 
Muon decay with rate $\lambda_+$ can proceed from all states. The small capture contributions from the upper
hfs states have been omitted.  }
\end{center}
\vspace{-.0cm}
\end{figure} 
  
\begin{figure}[t]
  \begin{center}
  \includegraphics[scale=0.5, angle=0]{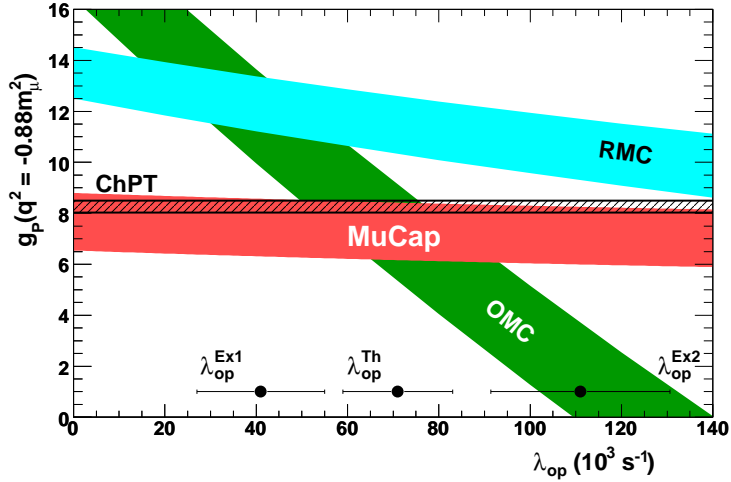}
  \vspace{0mm}
  \caption{Experimental and theoretical determinations of 
  ${\textsl g}^{}_P$, presented vs.\ the ortho--para transition 
  rate $\lambda_{\rm op}$ in the $p\mu p$ molecule.  The most 
  precise previous ordinary muon capture (OMC) experiment~\cite{Bardin:1980mi} and the RMC
  experiment~\cite{Wright:1998gi} both depend significantly on the 
  value of $\lambda_{\rm op}$, which itself is poorly known due to 
  mutually inconsistent experimental 
  ($\lambda^{\rm Ex1}_{\rm op}$~\cite{Bardin:1981cq},
  $\lambda^{\rm Ex2}_{\rm op}$~\cite{Clark:2005as}) and theoretical 
  ($\lambda^{\rm Th}_{\rm op}$~\cite{Bakalov:1980fm}) results.  
  In contrast, the MuCap result for ${\textsl g}^{}_P$ is nearly 
  independent of molecular effects.}
  \label{g_P.fig}
  \end{center}
\vspace{-8mm}
\end{figure}

However, in spite of efforts  spanning the last 40 years, the experimental situation remained inconclusive. 
Experiments lacked sufficient precision and could not be interpreted with confidence. 
A first measurement~\cite{Wright:1998gi} of radiative muon capture (RMC) on the proton suggested a value for $g_P$  
exceeding the chiral prediction by nearly 50\%.
Fig.~\ref{kin.fig} (top) illustrates the kinetic problem. The capture rate from $\mu p$ atoms and $p\mu p$ molecules 
are different combinations of the basic singlet capture rate \RS. In particular, in liquid hydrogen targets, $p\mu p$ molecules
are quickly formed proportional to $\phi \lambda_{pp}$ and the ortho-para conversion rate $\lambda_{op}$ 
is poorly known. $\phi$ is the hydrogen density normalized to LH$_2$. 
There has been a long-standing discrepancy between experiment~\cite{Bardin:1981cq} and 
theory~\cite{Bakalov:1980fm} on $\lambda_{op}$  and a more recent measurement~\cite{Clark:2005as} does not agree with 
either result. As a consequence, the situation prior to MuCap was inconclusive and exhibited mutually inconsistent 
theoretical predictions and experimental determinations of both ${\textsl g}^{}_P$ and $\lambda_{\rm op}$ 
(see Fig.~\ref{g_P.fig}). It is evident from this figure, that the previous experimental world average
of $g_P(exp)=10.5 \pm 1.8$, evaluated in~\cite{Gorringe:2002xx} assuming $\lambda_{op}$ from 
experiment~\cite{Bardin:1981cq} only, has
to be revised and its error significantly inflated if the full spread of $\lambda_{op}$ results is taken into account.       

The MuCap experiment has developed a novel technique  based on tracking the incoming muons in a time projection 
chamber (TPC) filled with ultra-pure deuterium-depleted hydrogen. This allows for a first precise measurement of muon capture 
in low-density gas, where $p \mu p$ formation is slow and 96\% of all captures proceed from the 
$\mu p$ singlet state.  The capture rate is determined from the difference between the measured disappearance rate 
$\lambda_- \approx \lambda_+ + \RS$ of negative muons in hydrogen and 
the $\mu^+$ decay rate~$\lambda_+$, where it is assumed that free $\mu^-$ and $\mu^+$ 
decay with identical rates according to the CPT theorem. 
An initial result~\cite{Andreev:2007wg} has just been released which clarifies the 
previously confusing landscape (Fig.~\ref{g_P.fig}). 
The capture rate from the hyperfine singlet ground state of the $\mu p$ atom is measured to be 
\RS\ = 725.0 $\pm$ 17.4~\ins, from which  
$g_P(q^2=-0.88\,m_\mu^2)=7.3\pm1.1$, is extracted. The result agrees 
within 1$\sigma$ with the EFT calculations and does not confirm the dramatic 
discrepancy to theory, which the RMC result had initially implied.

\begin{table}[ht]
\begin{center}
\begin{tabular}{lccc} \hline
\bf 			& \bf MuCap 2007 	&\bf MuCap Final&\bf MuSun	\\
 				& $\delta$\RS(\ins)    	& $\delta$\RS(\ins) 	&$\delta$\RD(\ins) \\	
\hline
\multicolumn{4}{l}{\bf Statistics}							\\	
				&	12.5	&	3.7	&	3.4		\\	
\hline
\multicolumn{4}{l}{\bf Systematics}							\\	
muon-induced kinetics		&	5.8	&	2	&0.5		\\  	
chemical impurities		&	5.0	&	2	&	2   	\\	
$\mu$d diffusion		&	1.6	&	0.5	&0.5	    	\\	
$\mu$p diffusion		&	0.5	&	0.5	&	    	\\	
\hline 	
\end{tabular}
\end{center}
\caption{Partial error budget for capture rates from published MuCap result (MuCap 2007), anticipated final result 
(MuCap Final) and planned MuSun experiment (MuSun), showing uncertainties due to muon atomic effects. }
\label{systematics.tab}
\end{table}

The MuCap experiment is ongoing, with improved systematics, ten times higher statistics, and
about three times reduced uncertainties expected for the final result. Table~\ref{systematics.tab}
compiles the muon atomic physics uncertainties affecting the final MuCap result on \RS.
Evidently, several subtle effects have to be accurately controlled to achieve the desired sub-percent precision.
According to Fig.~\ref{kin.fig} the corrections due to $p\mu p$ formation depend on the two rates $\lambda_{pp}$ and
$\lambda_{op}$. The molecule formation rate $\lambda_{pp}$ will be measured 
in a dedicated experiment with the MuCap detector. Neutron detectors have been installed to 
study the effect of  $\lambda_{op}$ on the capture neutron time distribution, but further theoretical
investigations should try to clarify the present conflicting situation and estimate the amount of 
density dependence of this rate between $\phi=1$ and the density of MuCap $\phi=0.01$.
Regarding the chemical impurities, a sophisticated continuous filtering system~\cite{Ganzha:2007uk}
 cleans the TPC gas to
ultra-high purity of $\approx$ 10~ppb. The purity is monitored {\em in-situ} during the experiment, by
observing capture recoils from the capture process $\mu + (A,Z) \rightarrow (A,Z-1) + \nu$ in the TPC, which
occurs after muons have been transferred from hydrogen atoms to trace impurities like N$_2$ and H$_2$O.
Corrections are applied to the measured lifetime after calibrating these processes with known 
concentrations of impurities, and new values for these transfer rates will be determined by MuCap. 
Finally, muonic hydrogen atoms can diffuse away from their point of origin and thus introduce 
a time dependent correction to the detector acceptance. While the $\mu p+p$ cross section 
is sufficiently large to keep $\mu p$ atoms within millimeters,
the famous Ramsauer-Townsend minimum in the $\mu d+p$ cross sections allows $\mu d$
diffusion over centimeter distances
at MuCap conditions. Since 2006, the protium gas used in the experiment is isotopically cleaned to
c$_D<$ 10 ppb with a custom built cryogenic separator. For the final analysis 
diffusion effects should be simulated for realistic experimental conditions, following the work 
in~\cite{adamczak:042718}.

\section{The MuSun Experiment}
\label{MuSun}

Muon capture on the deuteron, reaction (\ref{m+d.eq}), is the simplest weak interaction process on a nucleus 
which can both be calculated and measured to a high degree of precision. While the one-body contributions 
to this process are well defined by the elementary amplitudes of 
process~(\ref{m+p.eq}), the challenge lies in the
short-distance part of the axial two-body current. Traditionally, these effects have been modeled
with meson exchange currents~\cite{Tatara:1990eb,Adam:1990kf, Doi:1989kv}. During the last few years 
EFT calculations have developed a model-independent description, where the two-nucleon current 
is parameterized by a single low energy constant which integrates all the poorly constrained
short-distance physics. This constant is called $L_{1A}$ in the pion-less 
theory~\cite{Chen:2005ak} and $\hat{d}^R$ in ChPT~\cite{Ando:2001es}. 
Exactly the same constant dominates the theoretical uncertainty in fundamental weak 
astrophysics processes, like the $p+p \rightarrow d +e^+ + \nu_e$ reaction, 
which is the primary energy source in the sun and the main sequence stars, and the 
$\nu + d$ reaction, which provided convincing evidence for solar neutrino
oscillation, as both its charged current and neutral
current modes are observed simultaneously at the Sudbury Neutrino 
Observatory~\cite{Aharmim:2007nv}. While eventually one can hope to calculate this parameter on 
the lattice~\cite{Detmold:2004qn}, at the moment it has to be determined from experiment. Alas, 
existing experiments on the axial-vector interaction in the two-nucleon system are of 
limited precision~\cite{Butler:2002cw,Chen:2002pv}, so that one has to resort to the theoretically more 
complex three-nucleon system. A measurement of $\mu+d$ capture could fix $L_{1A}$,  
$\hat{d}^R$ within a fully consistent framework and could determine the solar pp fusion and 
$\nu d$ cross sections at essentially the same precision 
as the measured capture rate \RD. However, the best existing 
capture experiments~\cite{Bardin:1986,Cargnelli:1989} are not precise enough and the most
precise result~\cite{Bardin:1986} differs from modern 
theory by 2.9 standard deviations.

\begin{figure}[htb]
\begin{center}
\resizebox*{0.48\textwidth}{!}{\includegraphics{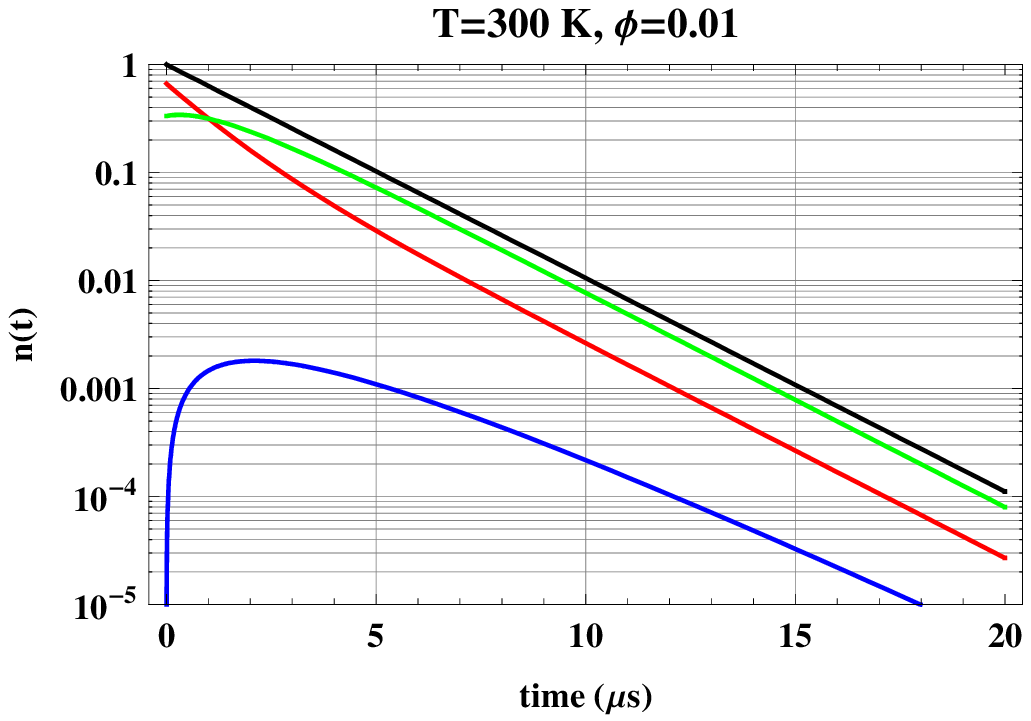}}  \resizebox*{0.48\textwidth}{!}{\includegraphics{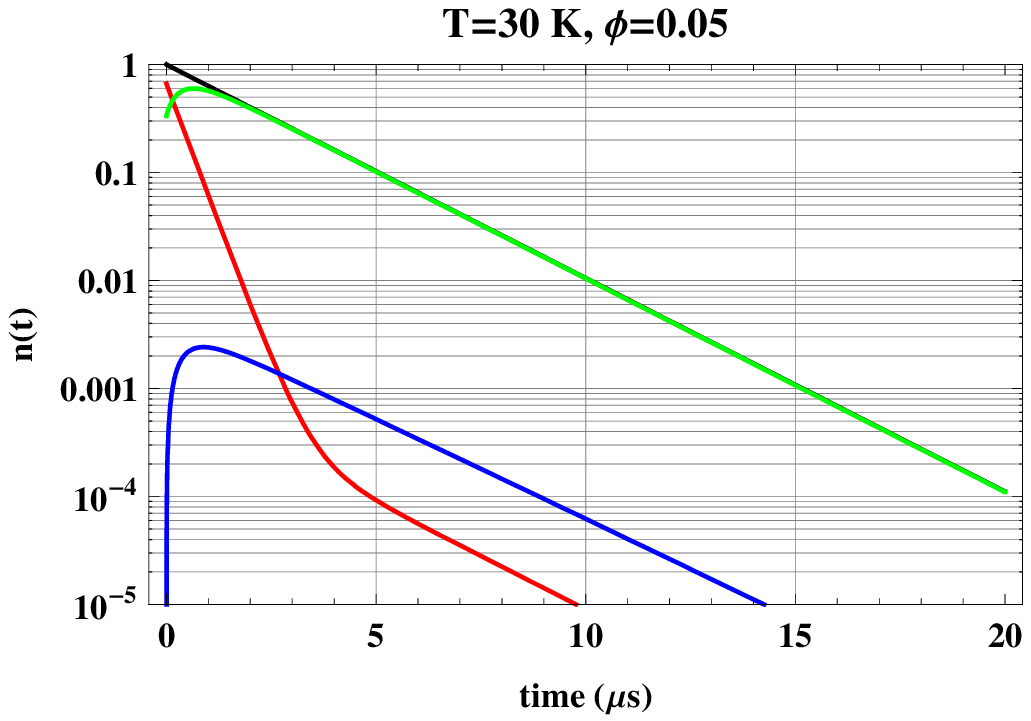}} 
\vspace{-.5cm}
\caption{Time distributions of relevant states (black=$\mu d$, green=$\mu d(\uparrow \downarrow)$, red=$\mu d(\uparrow \uparrow)$, blue=$\mu^3$He) for different deuterium densities
$\phi$ and temperatures T. The left panel indicates MuCap conditions, the right panel the optimized 
running conditions for the MuSun experiment. }
\label{time.fig}
\end{center}
\end{figure}

The new MuSun experiment~\cite{MuSun} will measure the capture rate \RD\ from the doublet state of the muonic
deuterium atom $\mu d$ to a precision of better than 1.5~\%. The measurement, based on
the MuCap techniques,  will have a significantly higher precision than previous experiments 
on $\mu +d$ capture and on other weak two-nucleon reactions, like $\nu +d$ scattering.
As seen in fig.~\ref{kin.fig} (bottom), for a clear interpretation and for the 
accumulation of sufficient capture statistics,  the target conditions should be chosen such that the 
$d\mu$ doublet state dominates and the formation of 
$\mu ^3$He is minimized. Although the kinetics is complicated, the
$d\mu$ system has been intensively studied as the prototype for resonant muon-catalyzed 
fusion~\cite{Breunlich:1989vg}. Our optimization shown in fig.~\ref{time.fig}  indicates that the 
target density should be increased to $\phi$=0.05, to accelerate the relatively 
slow hyperfine transition according to 
the rate $\phi \qdr$, where $\qdr = (37.0 \pm 4)~10^6~\ins$ at T = 30 K. The low temperature
is preferable to reduce $\dr$ to its non-resonant value, suppress upwards hfs transitions to the
quartet state and provide a large difference between doublet and quartet formation rates \dr\ and \qr,
respectively, which allows for  {\em in-situ} monitoring of the hfs kinetics by 
the observation of muon-catalyzed fusion reactions~\cite{Zmeskal:1990}.     
Table~\ref{systematics.tab} shows that the kinetic uncertainties to \RD\ can be reduced to an 
almost negligible  0.5~\ins\ 
at these conditions. However, the target purity requirements for the MuSun experiment are extremely stringent
(1 ppb) as the transfer rates to nitrogen scale with $\phi$ and are measured to be four times 
higher for $\mu d$ compared to $\mu p$ atoms~\cite{PhysRevA.57.1713}. Dedicated
experiments to remeasure these rates and to develop ultra-clean filtering and monitoring techniques
are foreseen. We also will study the residual muon polarization in $\mu d$, which might
be observable due to the relatively slow hyperfine transition rate \qdr.

\begin{figure}[tbh]
  \begin{center}
  \includegraphics[scale=0.45,angle=-90.]{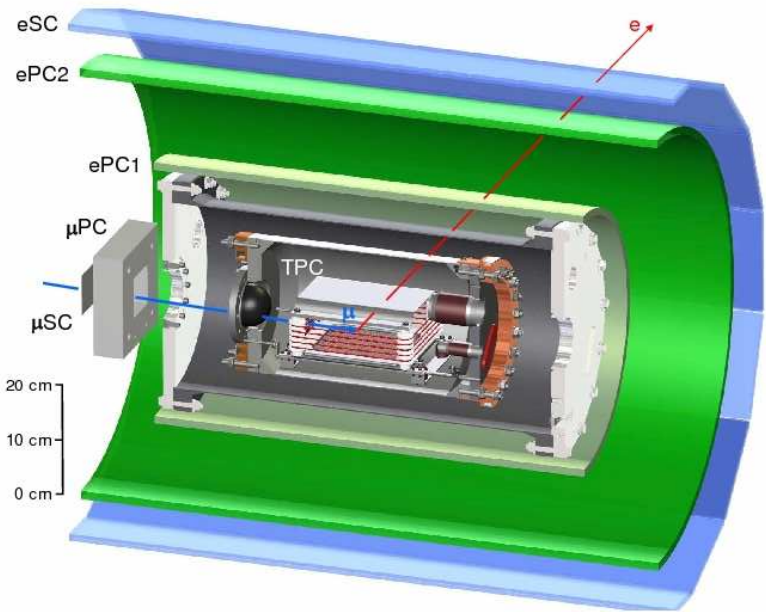}
  \caption{Simplified cross-sectional diagram of the MuSun detector.}
  \label{setup.fig}
  \end{center}
\vspace{-4mm}
\end{figure}

The schematic set-up of the MuSun experiment is shown in Fig.~\ref{setup.fig}, where a small high
density cryo-TPC filled with ultra-pure deuterium is embedded in an insulation vacuum vessel 
at the center of the electron tracking detectors.

\section{Positive Muon Lifetime Experiments}
\label{MuLan}
The Fermi constant $G_F$ is a fundamental constant of nature, which, together with $\alpha$ and $M_Z$,
defines the gauge couplings of the electroweak sector of the standard model.
It is directly related to the 
free muon decay rate, but, in the past, the extraction of  $G_F$ from  $\lambda_+$
was limited by unknown 2-loop radiative corrections. 
With those calculated~\cite{vanRitbergen:1998yd,vanRitbergen:1999fi}, 
the muon lifetime $\tau_\mu $, known to 18 ppm, became the limiting factor. Moreover, the precise knowledge
of the $\mu^+$ decay rate $\lambda_+ = \frac{1}{\tau_\mu}$ is required for the muon capture
experiments described above, as the capture rates are derived from the difference 
$\lambda_- -\lambda_+$.
  
The MuLan experiment is a measurement of $\tau_\mu $ to 1 ppm precision. 
A time-structured muon beam is generated by a fast electrostatic kicker 
and stopped in a target with internal or external magnetic field, to control the
initial muon polarization. During the beam off period positrons are recorded by a 
highly-segmented, symmetric detector.  A first publication~\cite{Chitwood:2007pa} on a limited data set  
gives $\tau_\mu$(MuLan) = 2.197013(24) $\mu$s.  The updated world average
$\tau_\mu$(World) = 2.197019(21) $\mu$s determines the Fermi constant $G_F$(World) = 1.166371(6) $\times$ 10$^{-5}$ GeV$^{-2}$ (5 ppm). Recently, the FAST experiment~\cite{Barczyk:2007hp} released a new
result $\tau_\mu$(FAST) = 2.197083(35) $\mu$s. The next major step is
to improve the precision to 1 ppm. MuLan has already collected the required two orders 
of magnitude higher statistics, which is currently being analyzed.


\end{document}